\documentclass[%
 aip,
 amsmath,amssymb,
superscriptaddress]{revtex4-1}

\usepackage{graphicx}
\usepackage{dcolumn}
\usepackage{bm}

\usepackage[utf8]{inputenc}
\usepackage[T1]{fontenc}
\usepackage[dvipsnames]{xcolor}
\usepackage{mathptmx}
\usepackage{amsmath}
\usepackage{etoolbox}
\usepackage{physics}
\usepackage{charter}
\usepackage{tempora}
\DeclareUnicodeCharacter{2014}{\sar}
\usepackage[printwatermark]{xwatermark}
\newwatermark[allpages,color=black!30,angle=45,scale=2,xpos=0,ypos=0]{Accepted Manuscript}
\makeatletter
\def\@email#1#2{%
 \endgroup
 \patchcmd{\titleblock@produce}
  {\frontmatter@RRAPformat}
  {\frontmatter@RRAPformat{\produce@RRAP{*#1\href{mailto:#2}{#2}}}\frontmatter@RRAPformat}
  {}{}
}%
\makeatother
\begin{document}


\title{Coalescence of polymeric sessile drops on a partially wettable substrate\footnote{\textbf{The following article has been accepted by Physics of Fluids (AIP Publishing). After it is published, it will be found at https://doi.org/10.1063/5.0073936.}}}
\author{Sarath Chandra Varma}
 \affiliation{%
$^1$Department of Mechanical Engineering, Indian Institute of Science, Bangalore, Karnataka-560012, India
}%

\author{Aniruddha Saha}%
\affiliation{%
$^1$Department of Mechanical Engineering, Indian Institute of Science, Bangalore, Karnataka-560012, India
}%
\author{Aloke Kumar*}
 \email{alokekumar@iisc.ac.in}
 \affiliation{%
$^1$Department of Mechanical Engineering, Indian Institute of Science, Bangalore, Karnataka-560012, India
}%


\begin{abstract}

Coalescence of sessile polymeric fluid drops on a partially wettable substrate exhibits a transition from inertial to viscoelastic regime at concentration ratio $c/c^*\sim1$. Our findings unveil that the temporal evolution of the growing bridge height follows a power-law behaviour $t^{b}$, such that the coefficient $b$ continuously decreases from 2/3 in the inertial regime ($c/c^*<1$) to an asymptotic value of 1/2 in the visco-elastic regime ($c/c^*>1)$. To account for fluid elasticity and characteristic time-scale in the viscoelastic regime, a modified thin film equation under lubrication approximation has been proposed using the linear Phan-Thien-Tanner constitutive equation. The temporal evolution of the droplet has been evaluated by solving the modified one-dimensional thin-film equation using a marching explicit scheme. The initial droplet shapes are obtained by re-sorting to energy minimization. A good agreement between numerical and experimental results is obtained. 
\end{abstract}

\maketitle

\section{Introduction}
The phenomenon of coalescence is of critical importance in various natural and industrial processes such as raindrop formation \cite{rain1}, atmospheric disturbances \cite{thunder}, coating and spraying \cite{spray,coating} and crystallisation \cite{liq-crys}. Coalescence phenomena for droplets is governed by the dynamics of liquid bridge formation and its growth, wherein its temporal growth bears the signature of the underlying governing equation. In this singular event, the liquid topology changes rapidly to form a single daughter droplet - the singularity occurs during the formation of the meniscus connecting the two drops at the onset of coalescence. In controlled experiments, such events are studied in various geometric configurations depending upon the orientation of the two drops, i.e. pendant-pendant, pendant-sessile and sessile-sessile. The growth of the neck radius, $r$, with the time, $t$, in pendant-sessile Newtonian droplets\cite{PRL2011} has two regimes depending on the viscosity of the fluid. The neck radius evolution has the scale of $r\sim t$ in the viscous regime and $r\sim t^{1/2}$ in the inertial regime. However, a recent study on polymeric droplets coalescence\cite{our} proposed a scaling of $r\sim t^{0.36}$, emphasizing the effect of elasticity on coalescence. The universal behaviour in both the Newtonian\cite{PNAS2019} and polymeric droplets\cite{our} is presented in the literature.

The coalescence of two sessile drops on a substrate is characterised by an initial rapid and a subsequent slow growth of bridge at the point of contact between the two drops. This changes the shape of the merged droplet from an initial oblong to roughly circular shape. The kinematics of the overall phenomenon is primarily governed by the growth of two geometric parameters with time: the bridge width, $r_m$, observed from the top view and bridge height, $h_b$, observed from the side view.
Even though the top view of the coalescing drops on a substrate looks similar to that of spherical drops, the bridge dynamics is very different due to the presence of viscous stresses in the former. However, similar to the coalescence of pendant-sessile droplets, the fluid viscosity affects the bridge growth in sessile-sessile drops coalescence as well. For an apparently high viscosity fluid, Ristenpart\cite{PRL2006} proposed the scale of $r_m\sim t^{1/2}$ for a wettable substrate. Another study on a partially wettable substrate by M.W.Lee\cite{lamgmuir2012} showed that the bridge width has slower growth, $r_m\sim t^{1/4}$, as compared to the scale proposed by Ristenpart. Alternatively, J.F Hernandez\cite{PRL2012} showed that the bridge height grows linearly with time and exhibits a self-similar dynamics for both the symmetric and asymmetric coalescence of droplets. In a contrasting study by M.W.Lee\cite{lamgmuir2012}, the scaling law proposed for the growth in bridge height is shown to have a dependence on contact angle, such that ${h_b}/{R_o}\sim{({t}/{t_c})}^\alpha$, where $R_o$ is half of the contact length and $t_c$ is the characteristic time. The values of $\alpha$ was seen to vary from 0.5 to 0.86 for different contact angles. For apparently inviscid fluids\cite{PRL2013,PRE2007}, the temporal evolution of the bridge height follows the scale of $h_b\sim t^{2/3}$ for contact angles below $90^\mathrm{o}$, contrasted by $h_b\sim t^{1/2}$ for a contact angle of $90^\mathrm{o}$ \cite{PRL2013,POF2013}. The power-law behaviour of the bridge evolution obtained from experiments has been validated with numerical simulations using lattice Boltzmann method\cite{POF18,latticeboltz}, multi-fluid modelling \cite{multifluid}, lattice BGK method \cite{latticebgk} and  volume of fluid method\cite{POF25}in inertio-capillary regimes\cite{POF11,POF13,POF26}. Similar studies of sessile drop coalescence were performed on miscible fluids of different surface tensions\cite{POF31,POF32} in which delay in coalescence was observed. A delay was also observed in coalescence of ethanol and water\cite{POF35}, where drops are self-propelled due to ethanol evaporation.

There exist several applications of drop coalescence in rheologically complex fluids such as emulsions \cite{51,53}, microfluidic applications \cite{48} and interfacial rheology \cite{55}. In fact, the presence of elasticity and characteristic time scale in the fluid has shown significant deviation in coalescence dynamics signature as compared to Newtonian fluids \cite{our}. Despite some work in this area, coalescence behavior of viscoelastic droplets still remains a sparsely studied area.

In this present study, we investigate the coalescence of two polymeric droplets on a substrate at early time instances. We experimentally and theoretically show the time dependence of bridge height, $h_b$, between the two merging droplets, each with viscosity, $\eta$, contact angle, $\theta$, contact length, $2R_o$, surface tension, $\sigma$, and relaxation time, $\lambda$. The experimental results are found to be consistent with the modified thin film equation under lubrication approximation obtained from linear Phan-Thien-Tanner (PTT) constitutive equation \cite{PTT1977,PTT1,PTT2}. Considering the droplets to be in thermodynamic equilibrium, the initial shape is obtained by employing the energy minimization technique. We used polyethylene oxide (PEO) as a representative fluid having a single contact angle to study the coalescence phenomenon.

\section{Materials and methods}
PEO solutions are prepared by adding appropriate quantity of PEO of molecular weight, $M_w= 5\times10^6$ g/mol (Sigma-Aldrich) in de-ionized (DI) water to get various concentrations $c$ (w/v) of 0.01\%, 0.02\%, 0.05\%, 0.061\%, 0.1\%, 0.2\%, 0.3\%, 0.4\%, 0.5\% and 0.6\%. All the solutions are stirred at 300 rotations per minute for 24 hours. Concentrations of PEO are chosen such that the solutions belong to three different regimes\cite{bird}. The solutions having concentration ratios $c/c^*<1$, $1<c/c^*<c_e/c^*$ and $c/c^*>c_e/c^*$ ($c^*$ being critical concentration and $c_e$ being entanglement concentration) are in dilute regime, semi-dilute unentangled regime, and semi-dilute entangled regime, respectively. The detailed calculations of $c^*$ and $c_e$ are described in Rheology section. 

Experiments were performed on the glass substrate (Blue star, India) of dimensions $75\times25\times1.4$5 mm. Preceding the experiments, the glass slides are cleansed with detergent and sonicated with acetone for 20 mins followed by sonication in water for 20 mins. Subsequently, the slides are dried in a hot air oven at 95$^\circ$C for 30 mins.

Surface tension $\sigma$ of the solutions are measured using optical contact angle measuring and
contour analysis systems (OCA25) instruments from Dataphysics $^{\textregistered}$ by the pendant drop method. These solutions have surface tension values of $0.063 \pm 0.02 \mathrm{~N} / \mathrm{m}$. Density of the solutions is found to be nearer to that of water with a value of $1000 \pm 50 \mathrm{~kg} / \mathrm{m}^{3}$ and is obtained through the
measurements of mass and volume. We have thus assumed the density of the polymeric solutions
to be $1000 \mathrm{~kg} / \mathrm{m}^{3}$ throughout the present work.

\begin{table}[hbt!]
\caption{\label{tab:table1}Rheological properties of the solutions and the power law index $b$.}
\begin{ruledtabular}
\begin{tabular}{ccccc}
$c$ (w/v)&Concentration ratio($c/c^*$)&$\eta_o$ (mPa.s)&$\lambda$ (ms)&Power law index $b$\\
\hline
0 & - & 1 & DI Water & 0.64 ~~~\\
0.01 & 0.16 & 1.3 & 1.5 & 0.6 ~~~\\
0.02 & 0.32 & 1.5 & 1.5 & 0.57 ~~~\\
0.05 & 0.82 & 2 & 1.5 & 0.56 ~~~\\
0.061 & 1 & 3 & 1.5 & 0.54 ~~~\\
0.1 & 1.6 & 6 & 2 & 0.51 ~~~\\
0.2 & 3.9 & 18 & 2.7 & 0.49 ~~~\\
0.3 & 5 & 46 & 3.5 & 0.51 ~~~\\
0.4 & 6.5 & 60 & 74 & 0.52 ~~~\\
0.5 & 8.2 & 200 & 115 & 0.51 ~~~\\
0.6 & 9.8 & 500 & 160 & 0.49 ~~~\\

\end{tabular}
\end{ruledtabular}
\end{table}

\section{Rheology}
\subsection{Critical concentration and relaxation time}
Critical concentration, $\displaystyle c^*$, of PEO is obtained by applying the Flory relation $\displaystyle c^*=1/[\eta]$ alongside the Mark-Houwink-Sakurada correlation\cite{43} $[\eta]=0.072M_w^{0.65}$, where $[\eta]$ is the intrinsic viscosity. This leads to a critical concentration value of $\displaystyle c^*=0.061\%$ w/v . Entanglement concentration is obtained as $c_e=0.366\%$ w/v by using the relation $\displaystyle c_e\approx 6c^*$\cite{63}. The critical concentration and entanglement concentration thus obtained are in good agreement with experimental values, shown in Fig. S1. Fig. S1 represents the variation of specific viscosity $\eta_{sp}$ with concentration.
The deviations in $\eta_{sp}$ observed at the concentrations of 0.05\% w/v and 0.3\% w/v mark the critical concentration and entanglement concentration, respectively.

The intrinsic characteristic time-scale $\lambda$ associated with the polymer chains is estimated using the Zimm model \cite{bird}.

\begin{equation}
  \lambda_z=\frac{1}{\zeta(3\nu)}\frac{[\eta]M_w\eta_s}{\mathrm{N_A k_B} T}   
\end{equation}

where, $\displaystyle \lambda_z$ is the Zimm relaxation time, $\eta_s$ is the solvent viscosity, $\mathrm{k_B}$ is the Boltzmann constant, $\mathrm{N_A}$ is the Avogadro number, $T$ is the absolute temperature and $\nu$ is fractal polymer dimension determined using the relation $a=3\nu -1$. Here, $a$ is the exponent of Mark-Houwink-Sakurada correlation. However, the relaxation time of the solutions in dilute $\lambda_D$, semi-dilute un-entangled $\displaystyle \lambda_{\mathrm{SUE}}$ and semi-dilute entangled $\displaystyle \lambda_{\mathrm{SE}}$ regimes solutions are calculated using the correlations; $\lambda_D=\lambda_z$, $\displaystyle \lambda_{\mathrm{SUE}}=\lambda_z\Big(\frac{c}{c^*}\Big)^{\frac{2-3\nu}{3\nu-1}}$ and $ \displaystyle \lambda_{\mathrm{SE}}=\lambda_z\Big(\frac{c}{c^*}\Big)^{\frac{3-3\nu}{3\nu-1}}$ 
\cite{64,65,66}. The relaxation times obtained for the chosen concentrations along with corresponding concentration ratios $c/c^*$ are listed in Table-1.

\begin{figure}[htb]
\includegraphics[width=1\textwidth]{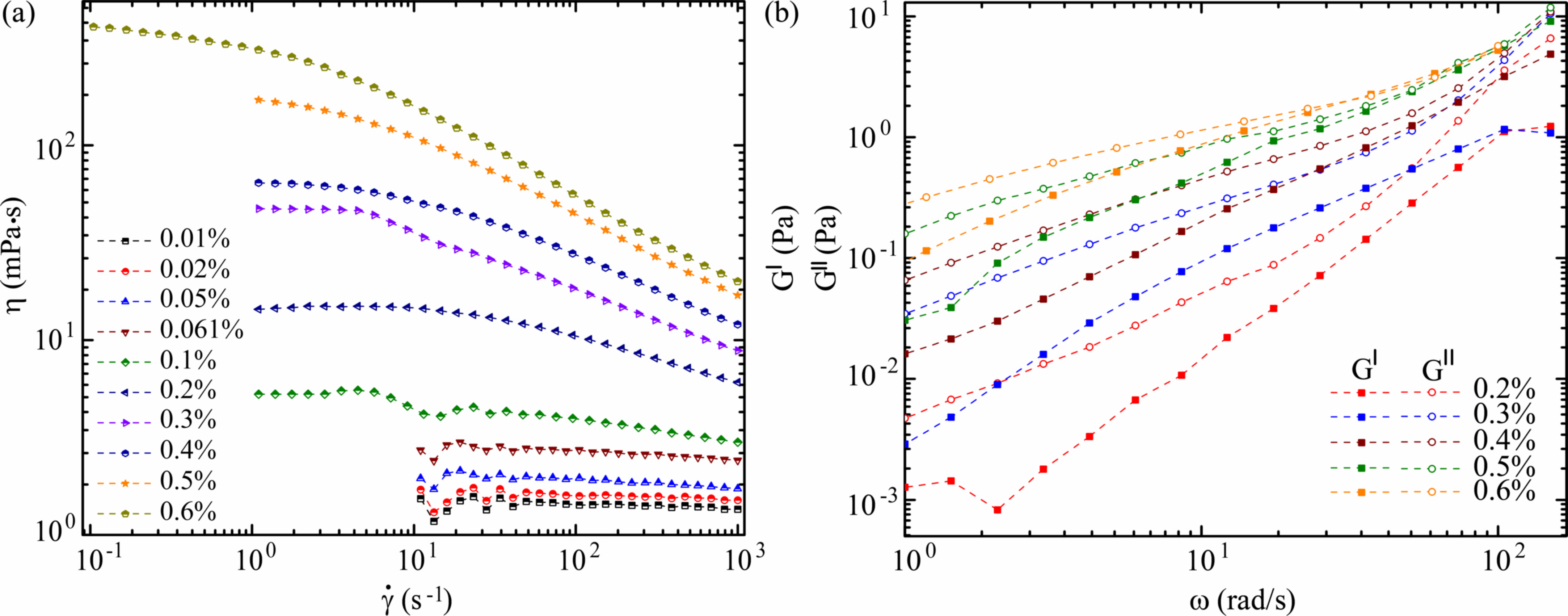} 
\caption{(a) Variation of viscosity with shear rate for different concentrations of PEO. (b) Frequency sweep representing change in storage modulus $G'$ and loss modulus $G''$ for 0.2\% w/v to 0.6\% w/v concentrations. }
\label{fig:rheology} 
\end{figure}

\subsection{Rheometry}
Viscoelastic behaviour of the solutions is characterized by Anton Paar$^{\tiny{\text{\textregistered}}}$ MCR 302 rheometer using a cone and plate 40 mm, 1$^{\circ}$ geometry. The variation of viscosity for the chosen solutions is shown in Fig.~\ref{fig:rheology}(a). Shear thinning behavior can be observed for all the concentrations. Zero shear viscosity, $\eta_o$, of the solutions listed in table -1 are obtained from the constant region of Fig.~\ref{fig:rheology}(a). To characterize the viscoelasticy of the polymer solutions, small amplitude oscillatory shear (SAOS) experiments are performed. The variation of storage modulus, $G'$ and loss modulus, $G''$, with frequency, $\displaystyle \omega$, is shown in Fig.~\ref{fig:rheology}(b) for 0.2\% w/v to 0.6\% w/v. A comparison between the values of the moduli, as shown in Fig.~\ref{fig:rheology}(b), suggests that the elastic forces are equally important as the viscous forces in governing the coalescence phenomenon. It must be noted here that the corresponding values of $G'$ and $G''$ are comparable even in the limits of frequency 1000 $\mathrm{s}^{-1}$ and 100 $\mathrm{s}^{-1}$, which occurs at 0.001 s and 0.01 s, respectively, and marks the lower and upper limit of our region of interest. The variation in either modulus is continuous and unlikely to shoot up from a smooth curve in the chosen span of 100 $\mathrm{s}^{-1}$ and 1000 $\mathrm{s}^{-1}$.

\section{Experiments}

\begin{figure*}[htb]
\includegraphics[scale=0.9]{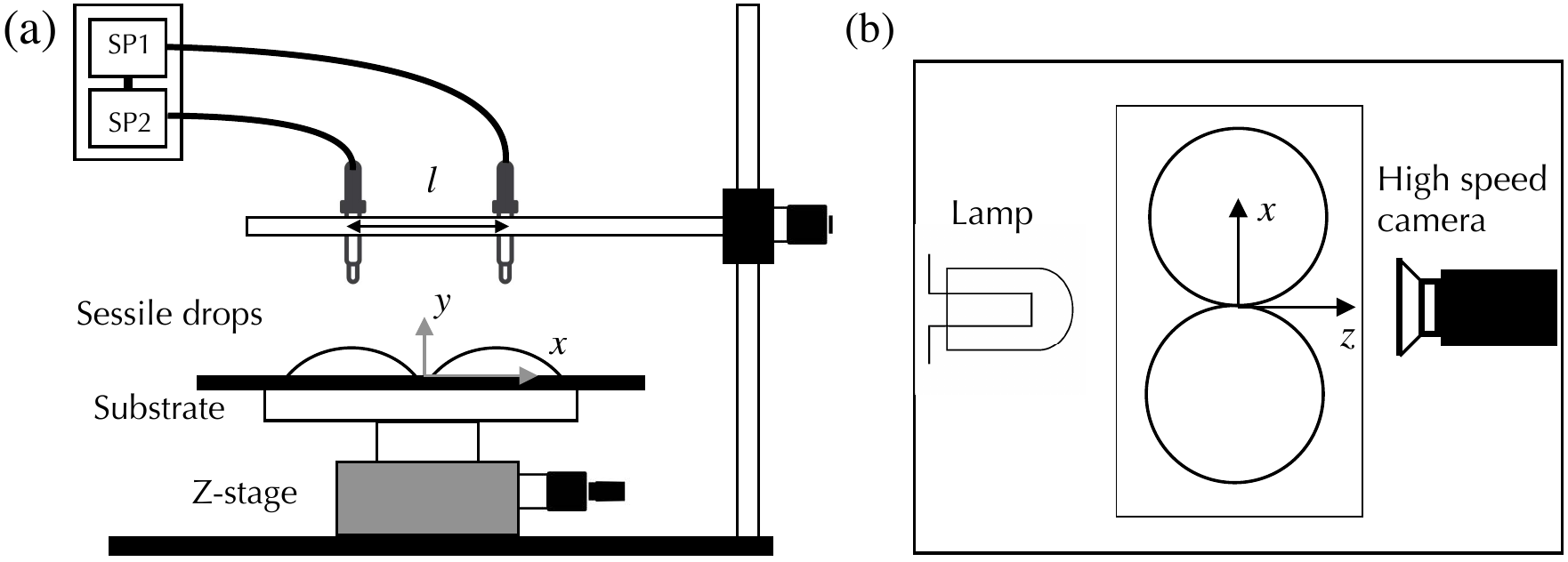} 
\caption{Schematic of the experimental setup (a)front view, (b)top view (SP1 and SP2 are syringe pumps)}
\label{fig:setup} 
\end{figure*}

Coalescence is achieved by creating two pendant drops of volume $\approx 11.5$ $\mu\mathrm{l}$ separated by a distance of $l\approx 5.5$ mm, using syringe pumps, as shown in Fig.~\ref{fig:setup}(a). Substrate is brought towards the pendant drops with an approach velocity $\sim 10^{-4}$ m/s. Once the droplets touch the surface, they begin to spread and the coalescence is set off instantaneously. Experiments are performed by illuminating the drops from behind using 45W LED light source (Nila Zaila, USA) at 100\% output as shown in Fig. 2(b). The process is recorded by Photron Fast-cam mini AX-100 high speed camera at 20,000 frames per second, using Navitar lens 6.5x zoom lens at a shutter speed of 1/90,000 s. A further analysis of the images is performed using custom algorithms in MATLAB. The background noise is removed by binarizing the images using an appropriate threshold. The change in pixel intensities along the column corresponding to bridge is tracked iteratively in each processed frame. The obtained coordinate values are then used to get the subpixel position of the bridge in original frames by considering neighboring gray intensities. Finally, when calculating the bridge height from the substrate, we subtract the obtained interface position from substrate position along the column corresponding to bridge. Contact angle of the drops is measured using the ImageJ DropSnake toolbox.

\begin{figure}[!h]
\includegraphics[width=0.6\textwidth]{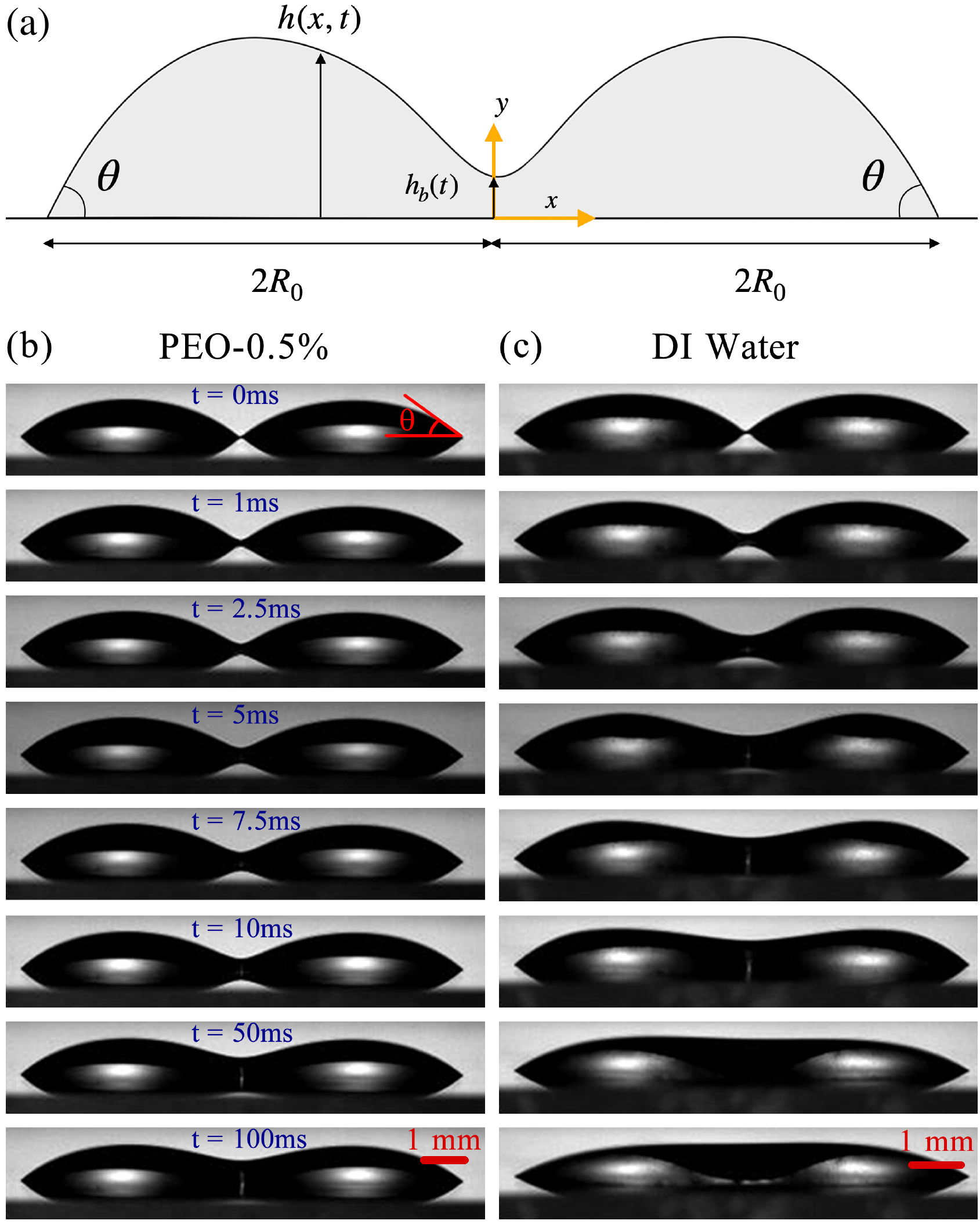} 
\caption{(a) Schematic of neck region during coalescence, (b) neck radius evolution of 0.5\% w/v concentration ($c/c^*=8.2$) of PEO on the left as compared to (c) DI water on the right at different instants.}
\label{fig:cascade} 
\end{figure}

As capillarity initiates the coalescence in sessile drops immediately after the drops come into contact, the phenomenon proceeds via the formation of a liquid bridge of an instantaneous height, $h(0,t)=h_b$, as depicted in Fig.~\ref{fig:cascade}(a). The kinematics of such formation is further guided by the local curvature effects and surface tension, $\sigma$. Such evolution of the liquid bridge during coalescence of two droplets of PEO with concentration ratio of $c/c^*=8.2$ and DI water are shown in Fig.~\ref{fig:cascade}(b) and (c) for different instants of time. The geometry is associated with a contact length of $2R_0\approx 5.25 \pm$0.25 mm and an interfacial contact angle of $\theta\approx35\pm3^{\circ}$ for all the solutions of PEO and DI water.

\section{Mathematical modelling}

\subsection{Modified thin film equation}
Coalescence phenomenon is captured theoretically by employing the framework of thin film lubrication approximation. The governing equations for the flow with the velocity field, $\va{v}(u,v,0)$, hydrostatic pressure, $p$, density, $\rho$, and extra stress tensor, $\bm{\tau}$ having the components $\tau_{xx},\tau_{xy}$ and $\tau_{yy}$ are given by conservation of mass, and momentum in $x$ and $y$ directions respectively, as shown in eq (2), eq (3) and eq (4), respectively.
\begin{equation}
\frac{\partial u}{\partial x}+\frac{\partial v}{\partial y}=0
\label{eq:vel1}
\end{equation}
\begin{equation}
\rho\left(\frac{\partial u}{\partial t}+u\frac{\partial u}{\partial x}+v\frac{\partial u}{\partial y}\right)=-\frac{\partial p}{\partial x}+\frac{\partial \tau_{xx}}{\partial x}+\frac{\partial \tau_{xy}}{\partial y}
\label{eq:vel2}
\end{equation}
\begin{equation}
\rho\left(\frac{\partial v}{\partial t}+u\frac{\partial v}{\partial x}+v\frac{\partial v}{\partial y}\right)=-\frac{\partial p}{\partial y}+\frac{\partial \tau_{yx}}{\partial x}+\frac{\partial \tau_{yy}}{\partial y}
\label{eq:vel3}
\end{equation}

By incorporating the lubrication approximations into eq (3) and eq (4) (See Supplementary information for the detailed non dimensionalization), they were reduced to eq (5). Governing equation for the bridge height evolution is derived by integrating the conservation of mass, which leads to eq (6).
\begin{equation}
\frac{\partial p}{\partial x}=\frac{\partial \tau_{xy}}{\partial y};\hspace{5mm}
\frac{\partial p}{\partial y}=0
\label{eq:vel4}
\end{equation}
\begin{equation}
\frac{\partial h}{\partial t}+u(h)\frac{\partial h}{\partial x}=-\int_{0}^{h}\frac{\partial u}{\partial x}dy.
\label{eq:vel5}
\end{equation}
Velocity profile of the flow along $x$-direction is determined by using the linear PTT constitutive relation as given in eq (7) where, $\bm{\tau}$ is the extra stress tensor, $\kappa$ is a model parameter and \textbf{\textit{D}} is the deformation rate tensor. 
\begin{equation}
\displaystyle \lambda\overset{\nabla}{\bm{\tau}}+{\bm{\tau}}\left[1+\frac{\kappa\lambda}{\eta_o}Tr(\bm{\tau})\right]=2\eta_o {\textbf{\textit{D}}}
\end{equation}

By assuming a steady state of stress in the fluid and introducing the lubrication approximations along with eq (5), the individual components of the stress tensor are obtained as:
\begin{equation}
\begin{aligned}
\tau_{xy}&=\frac{\partial p}{\partial x}(y-h) \\
\tau_{xx}&=\frac{2\lambda}{\eta_o}(\tau_{xy})^2 \\
\tau_{yy}&=0
\end{aligned}
\end{equation}
By substituting the individual components of stresses in eq (7), we get the velocity profile represented by eq (9).
\begin{gather}
\begin{split}
u=\frac{1}{2\eta_o}\frac{dp}{dx}\Big[y^2-2yh+\frac{2 \kappa\lambda^2}{\eta_o^{2}} \Big(\frac{dp}{dx}\Big)^2\Big(3y^2h^2  \\ -2yh^3-2y^3h+\frac{y^4}{2}\Big)\Big]
\label{eq:vel6}
\end{split}
\end{gather}
The velocity profile in eq (9) along with the pressure gradient terms expressed in terms of surface curvature, $\Gamma$ and gravity, $g$, as $\rho g \frac{\partial h}{\partial x}-\sigma \frac{\partial \Gamma}{\partial x}$, are substituted in eq (6) to obtain the final thin film equation as:

\begin{equation}
\frac{\partial h}{\partial t}=\frac{1}{3 \eta_o}\frac{\partial}{\partial x}\left[h^{3}\left(\rho g \frac{\partial h}{\partial x}-\sigma \frac{\partial^{3} h}{\partial x^{3}}\right)+\frac{6 \kappa \lambda^{2}}{5 \eta_o^{2}} h^{5}\left(\rho g \frac{\partial h}{\partial x}-\sigma \frac{\partial^{3} h}{\partial x^{3}}\right)^{3}\right]\\
\label{eq:BC}
\end{equation}
The required initial and boundary conditions for solving the thin-film equation are listed as:
\begin{equation}
\begin{aligned}
      h(x,0)&=h_0(x)\\
      \left.\frac{\partial h}{\partial x}\right|_{x=0}&=0 \\
      \left.\frac{\partial^3 h}{\partial x^3}\right|_{x=0}&=0\\
      \left. h\right|_{x=2R_0}&=0\\
      \left. \frac{\partial h}{\partial x}\right|_{x=2R_0}&=-\mathrm{tan}{\theta}\\
       \label{eq:thinfilm}
\end{aligned}
\end{equation}
\subsection{Initial droplet shape}
The initial droplet shape, $h_0(x)=h(x,0)$, is bilaterally symmetrical about the contact bridge that connects the two sessile drops. The free shape of the sessile drops prior to coalescence are determined by considering that they are in thermodynamic equilibrium and hence, occupy minimal free energy. Using the Young-Dupré equation, the contact angle $\theta$ can be expressed as 
\begin{equation}
\mathrm{cos}\theta=\frac{\sigma_{sg}-\sigma_{sl}}{\sigma_{lg}}
\end{equation}
where $\sigma_{sg},\sigma_{sl}$ and $\sigma_{lg}$ are surface tension values of the solid-gas,  liquid-solid and liquid-gas surface, respectively. The liquid drop is rotationally symmetric about a vertical axis which intersects the centre of the contact line. This allows us to neglect the shape variations in the azimuthal direction. The droplet shape is represented in $(R,\psi)$ coordinates where the origin coincides with the midpoint of the contact line, $\psi$ is taken clockwise for the present case. The thermodynamic potential, $G$ \cite{shape}, contains contributions from the bulk, $G_b$, the interfaces, $G_i$ and gravitational potential energy, $G_g$, for a constant gravitational field over the volume of the drop. Hence,
\begin{equation}
G=G_{b}+G_{i}+G_{g}    
\end{equation}
The thermodynamic potential, $G$, can be further expanded as:
\begin{equation}
\begin{split}
G=\left(\sigma_{lg}+\lambda_{l}\right) \left(\pi R/\sqrt{2}\right)^2 +\pi \int_{0}^{\pi / 2}\big[2 \sigma R \sqrt{\dot{R}^{2}+R^{2}}\\ +\frac{\rho g R^{4}}{2}  \cos \psi\big] \sin \psi \mathrm{d} \psi
\end{split}
\end{equation}
where, $\lambda_l$ is the partial free energy of the liquid-solid bonding per unit area of their contact interface \cite{searcy}. We have considered the volume of the droplet $V=2\pi/3 \int_{0}^{\pi/2}R^3 \mathrm{sin}\psi d\psi$ as a constraint and performed minimisation of the expression $G/\sigma-\mu V $, where $\mu$ is a Lagrange multiplier. We applied the Euler-Lagrange minimisation criterion to yield:
\begin{gather}
\Bigg[\frac{2 \dot{R}^{2}+R^{2}-\ddot{R} R}{(\dot{R}^{2}+R^{2})^{3 / 2}}+\frac{R-\dot{R} \cot \psi}{R \sqrt{\dot{R}^{2}+R^{2}}}\Bigg]+\frac{\rho g}{\sigma} R \cos \psi=\mu
\label{eq:shape}
\end{gather}
with the boundary conditions; $\dot{R}(0)=\mathrm{cos}\psi\sqrt{R(0)^2+\dot{R}(0)^2}$ and $\dot{R}(\pi/2)=0$. 

\subsection{Numerical solution}
The discretisation of the variable $h(x,t)$ is done using the explicit finite difference scheme\cite{gravity} where $h_i^n$ denotes the instantaneous height with $i$ as the spatial index and $n$ as the temporal index.  We have discretized the domain to $N_{x} \cdot N_{t}$ points that can be represented as $h_i^n\left(x_{i}, t_{n}\right) \mid i \in\left[1, N_{x}\right]$, $n \in\left[1, N_{t}\right]$. The iteration step is similarly denoted by $\Delta x=x_{i+1}-x_{i}$ and $\Delta t=t_{n+1}-t_{n}$, such that the points parallel to the $x$ -axis and $t$ -axis are evenly spaced, but $N_x$ and $N_t$ are not necessarily equal. For the spatial derivative, we have considered a differencing scheme with $2^{\mathrm{nd}}$ order accuracy \cite{finitediff}. The temporal derivative at the next time step $n+1$ is expressed in a discretized form with respect to the spatial derivatives at the current time step $n$ as:

\begin{equation}
\begin{split}
   \frac{h_i^{n+1}-h_i^{n}}{\Delta t}=\frac{1}{3\eta_o}\big[3(h_i^n)^2(h_i^n)'\big\{\rho g (h_i^n)'-\sigma(h_i^n)'''\big\}+(h_i^n)^3\big\{\rho g (h_i^n)''-\sigma(h_i^n)''''\big\}\\+\frac{6\kappa\lambda^2\sigma^2}{5\eta_o^{2}}\big\{5(h_i^n)^4(h_i^n)'\big\{\rho g(h_i^n)'-\sigma(h_i^n)'''\big\}^3+3(h_i^n)^5 \big\{\rho g(h_i^n)'-\sigma(h_i^n)'''\big\}^2\big\{\rho g(h_i^n)''-\sigma(h_i^n)''''\big\}\big\}\big] 
\label{eq:thinfilm}
\end{split}
\end{equation}

Eq~\eqref{eq:shape} is solved using shooting method with 4th order Runge-Kutta formulation. The model parameter $\kappa$ in eq~\eqref{eq:thinfilm} is used as the tuning parameter to match the experimental and computational results. The initial shape $h_0(x)=h(x,0)$ thus becomes $R(\psi)$, where $R\mathrm{cos}\psi=x$. For modelling of the coalescence, we have considered the conjunction of two adjacent shapes $h_0(x)$. The coalescence proceeds with $h(0,t\rightarrow{0^+})=h_y$, where $h_y$ is a sufficient small value denoting initial drop contact height. The evolution of the drop starts with the initial condition and the boundary conditions represented in eq (11). Explicit time-marching schemes are used by considering time steps that maintain numerical stability to obtain the solution of eq (10). 

\section{Results and Discussion}
Once the droplets touch the substrate, they begin to spread in order to attain thermodynamic equilibrium. In this process, they come in contact with each other and coalescence onsets with the formation of the singularity. The properties of the partially wettable substrate are reflected in the contact length, $2R_o$ and contact angle, $\theta$. These parameters of the drop-substrate are used to determine the energy minimized configuration. As the contact angle of the drops can be represented by eq (12), it is evident that the spreading parameter, $S$, is negative, and the substrate is partially wettable. The wettability of the substrate affects the bridge growth during the coalescence process. The bridge grows due to mass flux into the bridge from its close neighborhood at early instants of time. Though the equilibrium at the pinned ends may get disturbed during the coalescence process leading to change in contact angle, the time taken for the same is much greater than the time of interest of the present study. As a result, the contact angle at the pinned end can be safely considered to be constant without any loss of generality. The contact angle for all the solutions used in the present study is $35\pm3^\circ$, suggesting that the wetting property of the substrate is independent of concentration ratio. This constant contact angle results from constant surface tension values of all the solutions. The variability of the substrate is neglected as the contact angle and contact length of the drops are within the error bounds for all the trials, indicating that the material and geometric variability in the substrates are minimal.

\begin{figure}[h]
\includegraphics[scale=0.38]{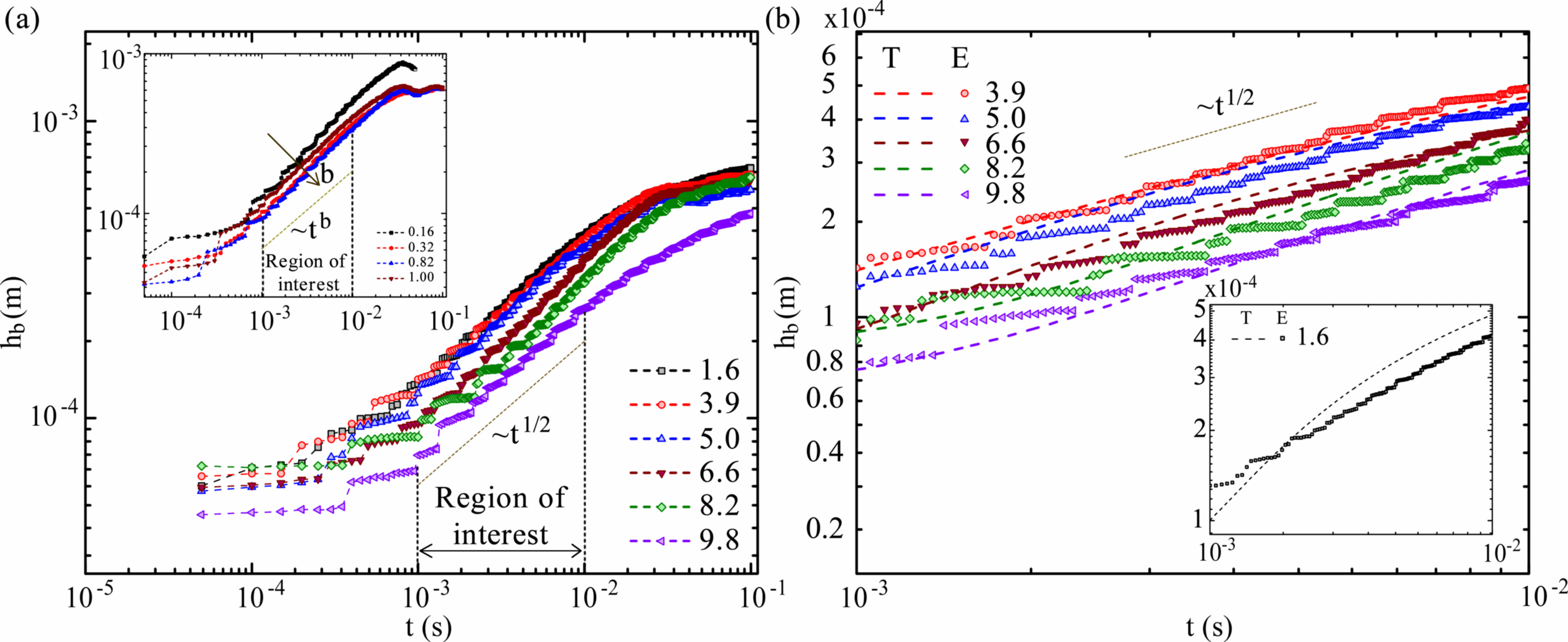} 
\caption{(a) Temporal evolution of bridge height for various concentration ratios of $c/c^*>1$ with an inset showing the bridge growth for $c/c^*<1$, (b) Comparison of bridge height evolution between experiments and numerical solution of eq (10) for various $c/c^*>1$ values with an inset showing the mismatch between the experiments and numerical solution for $c/c^*=1.6$. (Note: T represents theoretical and E represents experimental).
}
\label{fig:hvst} 
\end{figure}

The temporal variation of the bridge height, $h_b=h(0,t)$, for various concentration ratio, $c/c^*$, of the polymeric drops is shown in  Fig.~\ref{fig:hvst}(a). The bridge height evolution for all the concentrations represented in Fig. 4(a) are the averaged values of 5 trials. The error in the measurements is less than $\pm5\%$. It can be seen that following a period of power law growth, the bridge height reaches a stable value. We limit our attention to the region of power law growth, equivalently the linear region in Fig.~\ref{fig:hvst}(a). Inset of Fig.~\ref{fig:hvst}(a) shows that in the limit of $c/c^*<1$, an increase in polymeric concentration increases the deviation of $b$ from its initial value of 2/3 at $c/c^*=0$. It can also be observed from the same figure that the deviation subsequently reaches a stable value of 1/2 at $c/c^*\sim1$, with no decrease in rate of bridge evolution with further increase in polymer concentration. The values of $b$ for all the chosen solutions are given in table-1. The error in $b$ is recorded as $\pm5\%$. For the solution of $c/c^*>10$, the chains are highly entangled, which leads to delay in merging even after the formation of bridge, as shown in Fig. S3. In present study, we have only considered solutions of concentration ratio $c/c^*<10$. The bridge height represented in Fig.~\ref{fig:hvst}(a) is validated numerically by following the proposed numerical scheme, as expressed by eq (16). The bridge height evolution represented by Fig.~\ref{fig:hvst}(b) shows the agreement between the experimental and computational results. However, a discrepancy between experimental and computational results for $c/c*=1.6$ is highlighted in inset of Fig.~\ref{fig:hvst}(b). The reason for the discrepancy can be contributed to dropping out of the inertial terms in eq (3) along with the stress contribution due to  solvent viscosity in eq (7). Through Fig.~\ref{fig:hvst}(b), we have mathematically demonstrated the presence of a temporal scale of $t^{1/2}$ for polymeric fluid coalescence for the concentration ratio $c/c^*>1$.

\begin{figure}[h]
\includegraphics[scale=0.5]{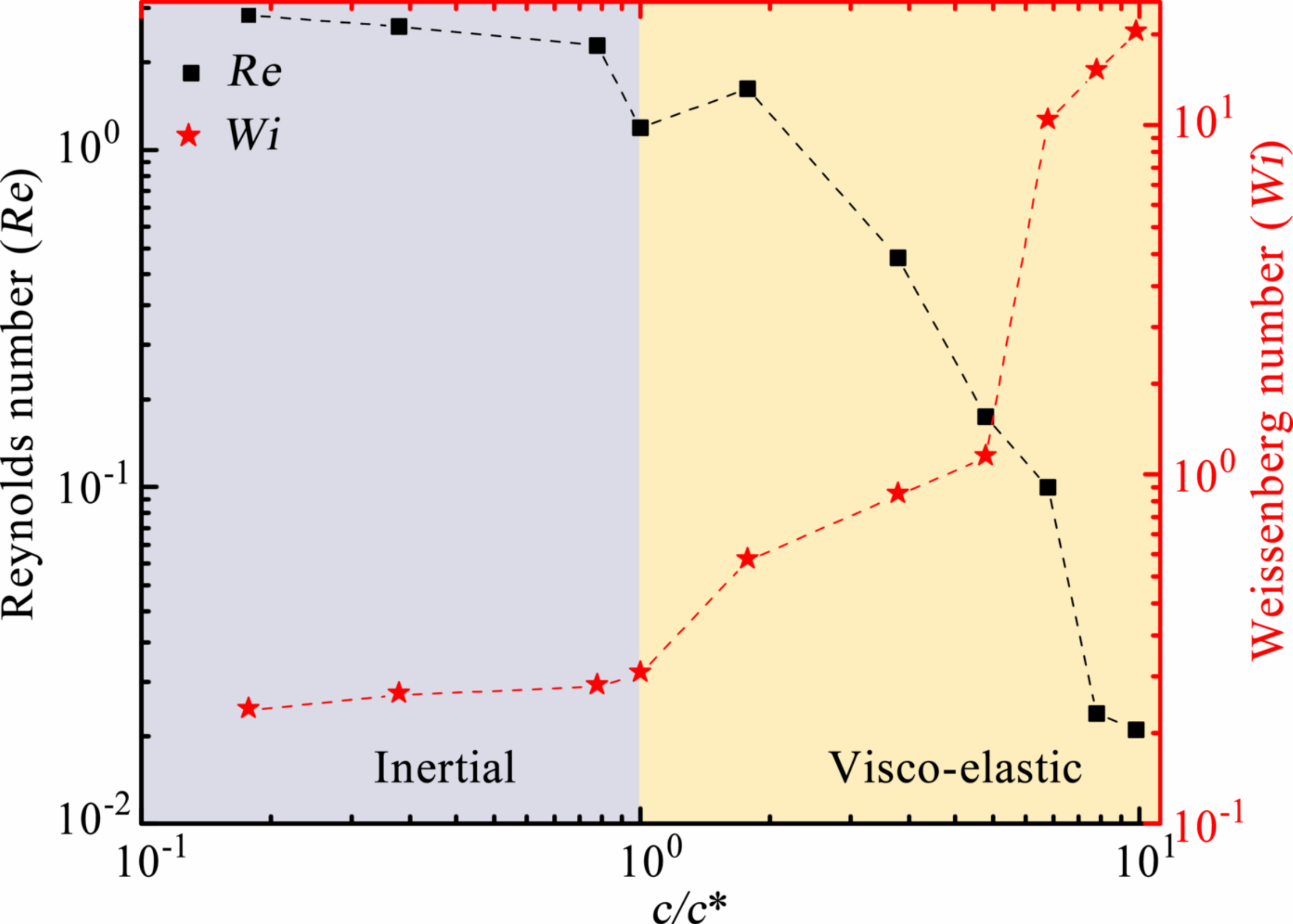} 
\caption{Comparison of predominant forces using Reynolds number and Weissenberg number for different concentration ratios.}
\label{fig:numbers} 
\end{figure}

As the droplets touch the substrate, the polymer chain tries to elongate along the shear direction. In the initial instances, when the drops begin to merge, the polymer chains are highly elongated. These elongated chains try to inhibit the coalescence, leading to deviation of the exponent from Newtonian behaviour. To capture this deviation in polymeric droplets, it is necessary to delineate the underlying forces governing the process. For polymeric droplets, it can be expected that the dominant forces will be delineated by two time averaged non-dimensional numbers - Reynolds number $Re=<\rho u_cl_c/\eta_o>$, and the Weissenberg number $Wi=<\lambda u_c/l_c>$, where $\eta_o$, $u_c$ and $l_c$ represent the zero shear viscosity, characteristic velocity and length scales, respectively, in the region of interest. The characteristic scales associated with the flow are $u_c\sim\partial h_b/\partial t$ and $l_c\sim h_b$. The values of $Re$ and $Wi$ obtained using these characteristics scales as represented in Fig.~\ref{fig:numbers}, show the transition from inertial regime to visco-elastic regime at $c/c^*\sim1$, except  for $c/c^*=1.6$. This suggests that the solutions of $c/c^*<1$ and $c/c^*>1$ are in inertial and visco-elastic regime, respectively. It is also observed from Fig.~\ref{fig:numbers} that $Wi\sim\mathcal{O}(1)$ for concentrations $c/c^*<c_e/c^*$, which suggests that the elastic forces and viscous forces are comparable to each other. For the solutions of $c/c^*=1$ and $c/c^*=1.6$, the Reynolds number $Re\approx1$ suggests that both the inertial and viscous forces are comparable. This may represent a transition from inertial to viscoelastic regime. However, the analysis of this transition regime is beyond the scope of current study. On the contrary, $Wi\sim\mathcal{O}(10)$ for concentration ratios $c/c^*>c_e/c^*$, which signifies the dominance of elastic forces over viscous forces. This increase in order of  $Wi$ for $c/c^*>c_e/c^*$ leads to a decrease in intercept of the temporal bridge height, as observed in Fig.~\ref{fig:hvst}(a). Therefore, a viscoelastic dominant behaviour has been considered throughout the analysis for $c/c^*>1$. The presence of elastic forces shown through $Wi$ leads to a modified velocity profile eq (9), which is further used to derive eq (10). It needs to be noted here that the proposed modified thin-film equation eq (10) cannot be used for the analysis of bridge growth for solutions in inertial regime, as the inertial terms are rendered non-dominant by the lubrication approximation.

\begin{figure*}[h]
\includegraphics[scale=0.70]{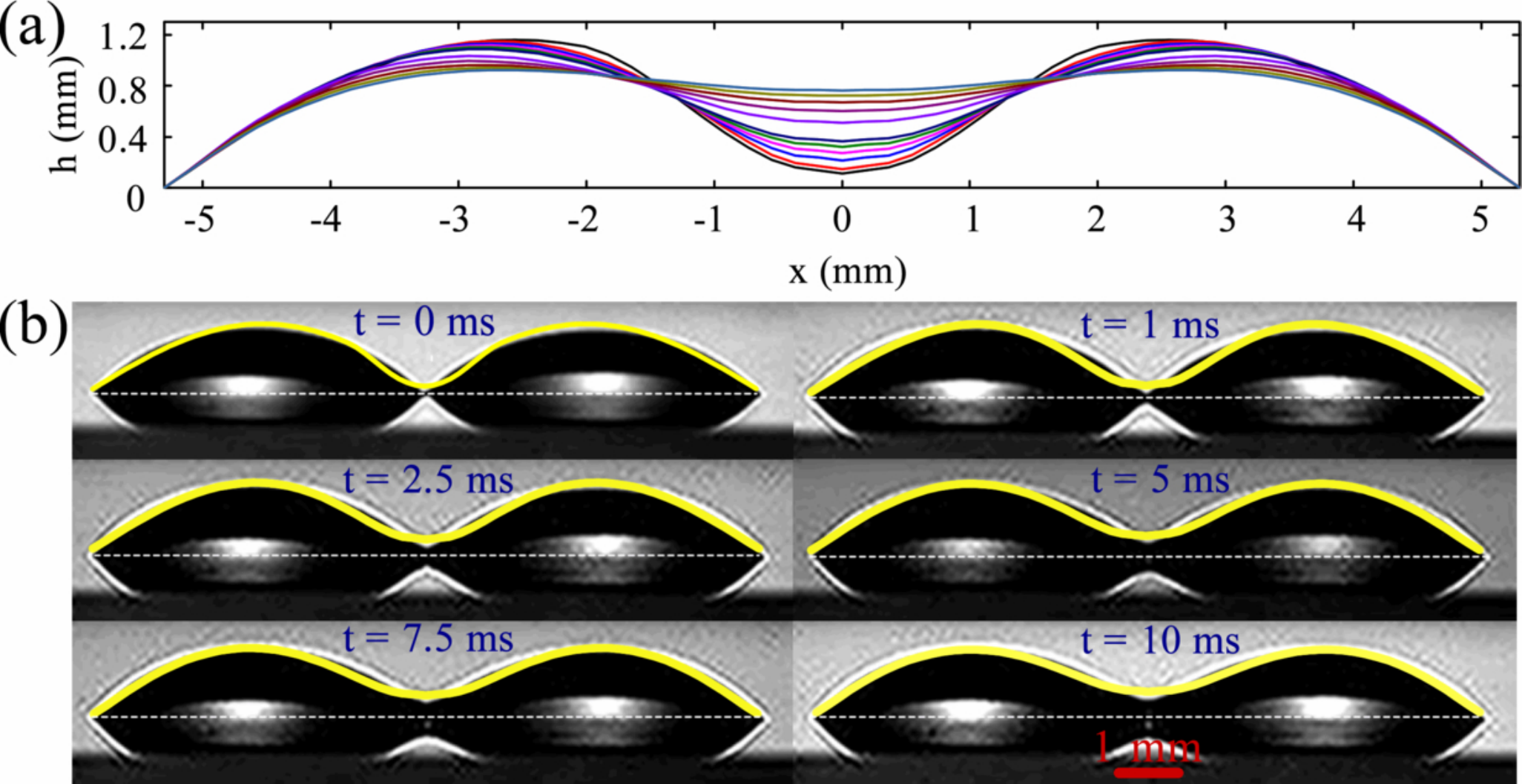} 
\caption{(a) Evolution of shape of the merging droplets during the coalescence process at time instances of 0ms, 2ms, 4ms, 6ms, 8ms, 10ms, 20ms, 30ms, 40ms, 50ms and 60ms for $c/c^*=8.2$. (b) Agreement between the computational and the experimental drop profiles of $c/c^*=8.2$ shown for six time instances.}
\label{fig:computational} 
\end{figure*}

The changes in the droplet shapes with the progress of coalescence is obtained from the solution of eq (10) and is represented in Fig.~\ref{fig:computational}(a) for $c/c^*=8.2$. The additional term in eq (10) that makes it a $3^{\mathrm{rd}}$ degree differential equation distinguishes it from the thin film equation of Newtonian fluid, which is a special case of $\lambda=\kappa=0$. The presence of this term makes a significant contribution to the evolution of the droplet shape during coalescence of polymeric fluids, as shown in Fig.~\ref{fig:computational}. However, the small changes in the spread-length observed experimentally are not taken into account in the computation, as the ends of the simulation domain have been kept constant. The agreement between the experimental and computational profiles of the drop shapes is represented in Fig.~\ref{fig:computational}(b) for $c/c^*=8.2$ at six time instances. We have also performed the analysis for the fluids used by Min Wook Lee et.al.\cite{lamgmuir2012} at a contact angle of $27^{\circ}$ by dropping the third-degree term and the gravity term in eq (10). The result obtained is in good agreement with the reported experimental values \cite{lamgmuir2012}, as represented in Fig. S2(a) (See Supplementary information). The evolution of the bridge growth for DI water, which is the base solvent to the solutions, is represented in Fig. S2(b) (See Supplementary information). However, the evolution of the liquid bridge cannot be captured by eq (10) by taking $\lambda=0$, as the coalescence of DI water takes place in the inertial regime\cite{PRL2013} where, Reynolds number $Re>1$.

\begin{figure*}[h!]
\includegraphics[scale=0.36]{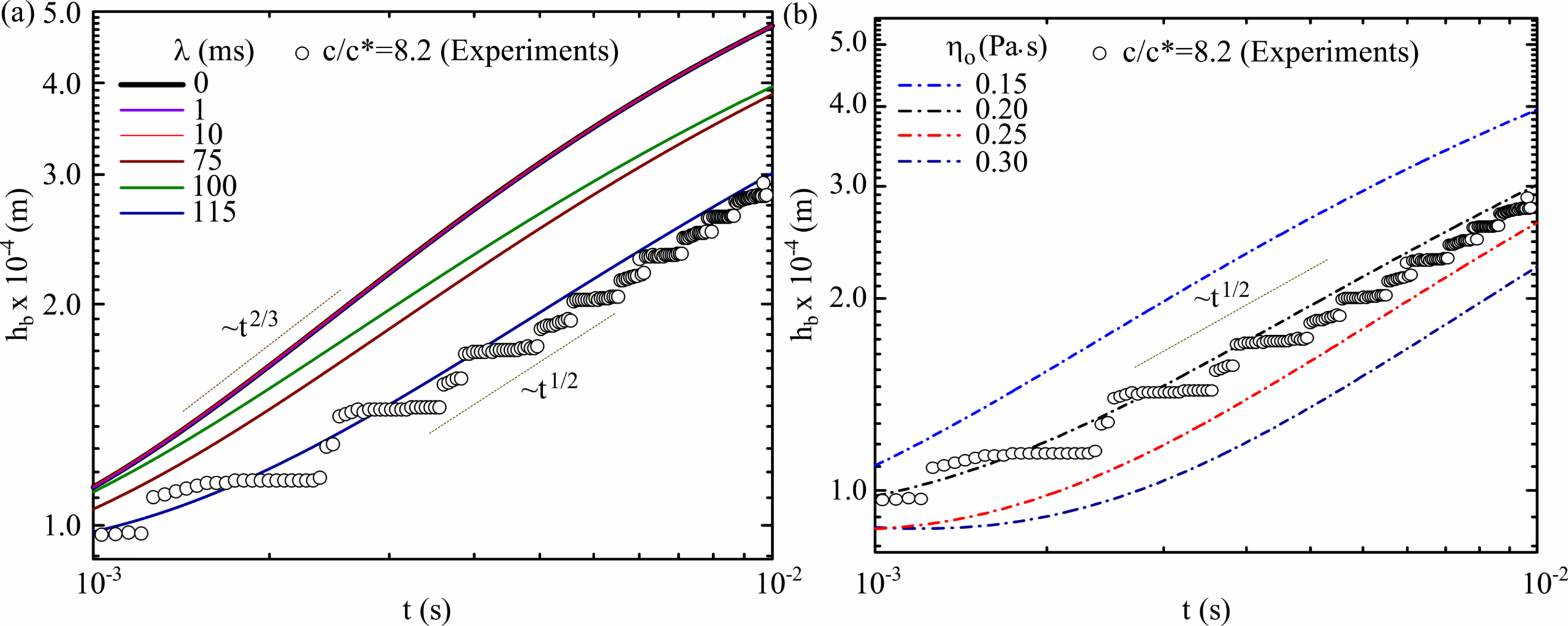} 
\caption{(a) Effect of relaxation time of the fluid on the bridge growth with constant viscosity $\eta_o=0.2$ Pa.s, (b) Bridge evolution due to variation of viscosity with constant relaxation time $\lambda=115$ ms.}
\label{fig:parametric} 
\end{figure*}

The subsequent discussion highlights the quantitative effect of the third-degree term having the coefficient $\frac{2\kappa\lambda^2}{5\eta_o^2}$ in eq (10) and establishes that $\lambda$ is the most important nodal parameters of a polymeric liquid, on which the power law behavior depends. Fig.~\ref{fig:parametric}(a) shows the variation of bridge height with time at different values of $\lambda$ keeping $\eta_o=0.2$ Pa.s and $\kappa=0.0005$. For the chosen conditions, the viscous time scale represented by the expression $\tau_v=\frac{2\eta_o R_o}{\sigma}$ turns out to be 15 ms. It is observed that when $\lambda>15$ ms, the slope of the curve attains a value of 1/2. On the contrary, when $\lambda<15$ ms, a visible shift in slope of the curve towards a value of 2/3 can be observed. The dynamics governing the above phenomenon lies in the complex interaction between polymer chains with respect to the viscous time scale. When the relaxation time scale of the polymer chains is lesser than the viscous time scale, the Newtonian behavior prevails. However, when relaxation time is larger than the viscous time scale, the polymeric chains persist in an unrelaxed state, causing relatively slow bridge growth. Similarly, variation of viscosity for a relatively high relaxation time $\lambda=115$ ms and $\kappa=0.0005$ as represented in Fig.~\ref{fig:parametric}(b), shows that chains are not relaxed irrespective of the chosen values of $\eta_o$, as the viscous time scale is always less than the relaxation time. It is thus the presence of an inherent time scale in polymeric fluid, whose relative magnitudes as compared to the viscous time scale affects the final nature of the coalescence phenomenon.

\section{Conclusion}
The current study presented for sessile drop coalescence involving polymeric fluids opens up a new perspective for understanding the behavior of non-Newtonian fluids. We have performed the experiments in the frontal plane to capture the temporal evolution of the bridge. The present study has been carried out over a wide range of polymer concentration ratios. It is observed that the transition from the inertial  regime to visco-elastic regime occurs at $c/c^*\sim1$. Theoretically, we have been able to demonstrate the accurate description of the coalescence through a modified thin-film equation in visco-elastic regime $\left(c/c^*>1\right)$, which can be further extended for a wider class of non-Newtonian fluids. The fluid-substrate interaction holds key insights to the pinning of drop during the process. The surrounding fluid effect can be an important factor behind the dynamics, which has been presently neglected by considering air as the outer fluid. This may require further exploration of the domain of applicability which might culminate into explaining the merger phenomenon between two fluid ensembles with characteristic scales ranging much higher than the values considered in the present study. Future work should be dedicated to a variety of cases, such as bio-fluids, colloids and also to unify the physics of this evolutionary phenomena. \\ 
\section*{Supplementary Material}
See supplementary information for the Newtonian droplets bridge growth during coalescence.
\begin{acknowledgments}
AK acknowledges partial support from DST-SERB grant no. EMR/2017/003025.
\end{acknowledgments}

\section*{Data Availability Statement}
The data that supports the findings of this study are available within the article [and its supplementary material].

\bibliography{bibfile}

\end{document}